# Regularities in the dynamics and development of the International System

*Ingo Piepers*

*Version: 23 October 2014*

*ingopiepers@gmail.com*
*global4cast.wordpress.com*

### Abstract

A finite-time singularity accompanied by log-periodic oscillations shaped the war dynamics and development of the International System during the period 1495 - 1945. The identification of this singularity provides us with a perspective to penetrate and decode the dynamics of the International System. Various regularities in the dynamics of the International System can be identified. These regularities are remarkably consistent, and can be attributed to the connectivity and the growth of connectivity of the International System.

## 1. Introduction

In this research report I present a number of regularities I have identified in the dynamics and development of the International System during the period 1495 - 1945. The recognition of these regularities is made possible by the identification of a finite-time singularity, accompanied by log-periodic oscillations, in the war dynamics of the International System. After presenting these regularities, I determine the statistical relationships between the variables involved, and argue that connectivity and connectivity growth of the International System provide a plausible causality for these statistical relationships and the observed regularities (see also: "*War: Origins and Effects, How connectivity shapes the war dynamics and development of the International System*" (2, 2014)). This analysis is based on the data set of Levy (1, 1983).

## 2. Regularities

In this section I present the regularities I have identified in the (war) dynamics and development of the International System.

### a. A finite-time singularity accompanied by log-periodic oscillations.

Below calculations are based on the identification of a finite-time singularity, accompanied by log-periodic oscillations in the war dynamics of the International System.
The identification of this singularity makes it possible to identify two types of wars: *systemic* and *non-systemic* wars.
Four systemic wars make up the singularity: (1) the Thirty Years' War, (2) the French Revolutionary and Napoleonic Wars, (3) the First World War, and (4) the Second World War.
At the critical time (t(c) = 1939) of the finite-time singularity the connectivity of the International System reached a critical threshold, and produced a critical transition.

At that stage, the high degree of connectivity of the International System and further drive for cooperation had become incompatible with the anarchistic nature of the International System.

The critical transition had two effects: (1) it transformed the 'European System' from an anarchistic system into a cooperative security community, and (2) it marked the start of the actual 'globalization' of the International System.

| Development of life spans of successive oscillations | | | |
|---|---|---|---|
| **Systemic War** | | **Time** | **t(c) – t** |
| Second World War | t(c) = t-critical | 1939 | 0 |
| First World War | t(1) | 1914 | 25 |
| French Revolutionary and Napoleonic Wars | t(2) | 1792 | 147 |
| Thirty Years' War | t(3) | 1618 | 321 |

*Table 1. This table denominates the systemic wars and their 'timings' making up a finite-time singularity accompanied by three log-periodic oscillations.*

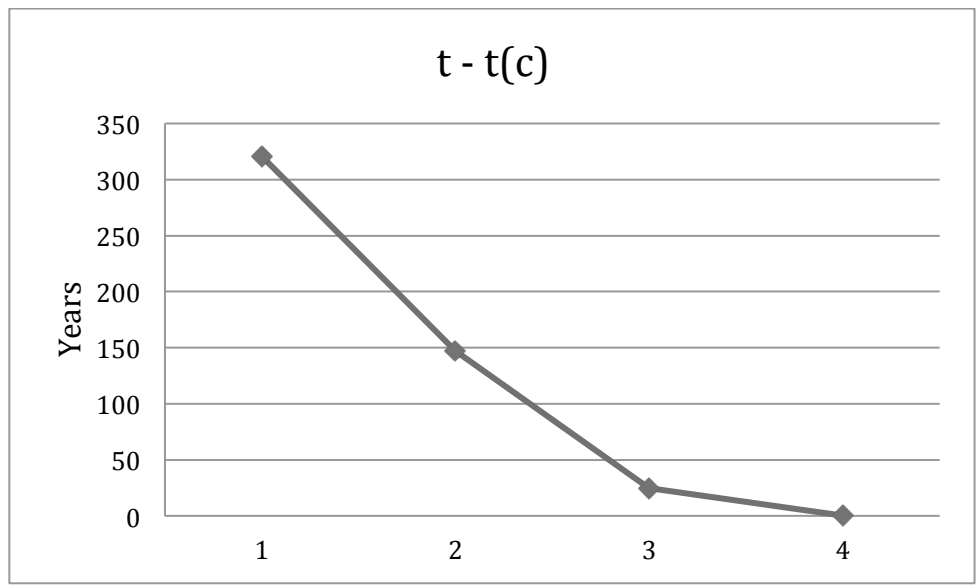

*Figure 1: Identification of a finite-time singularity accompanied by log-periodic oscillations in the war dynamics of the International System.*

The mathematical representation of this singularity is: ***Life span(t) = 19.6e^(0.936 t), with R2 = 0.99***. Because these oscillations are periodic in the logarithm of the variable (K(c) – K) / K(c), we refer to them as 'log-periodic.'

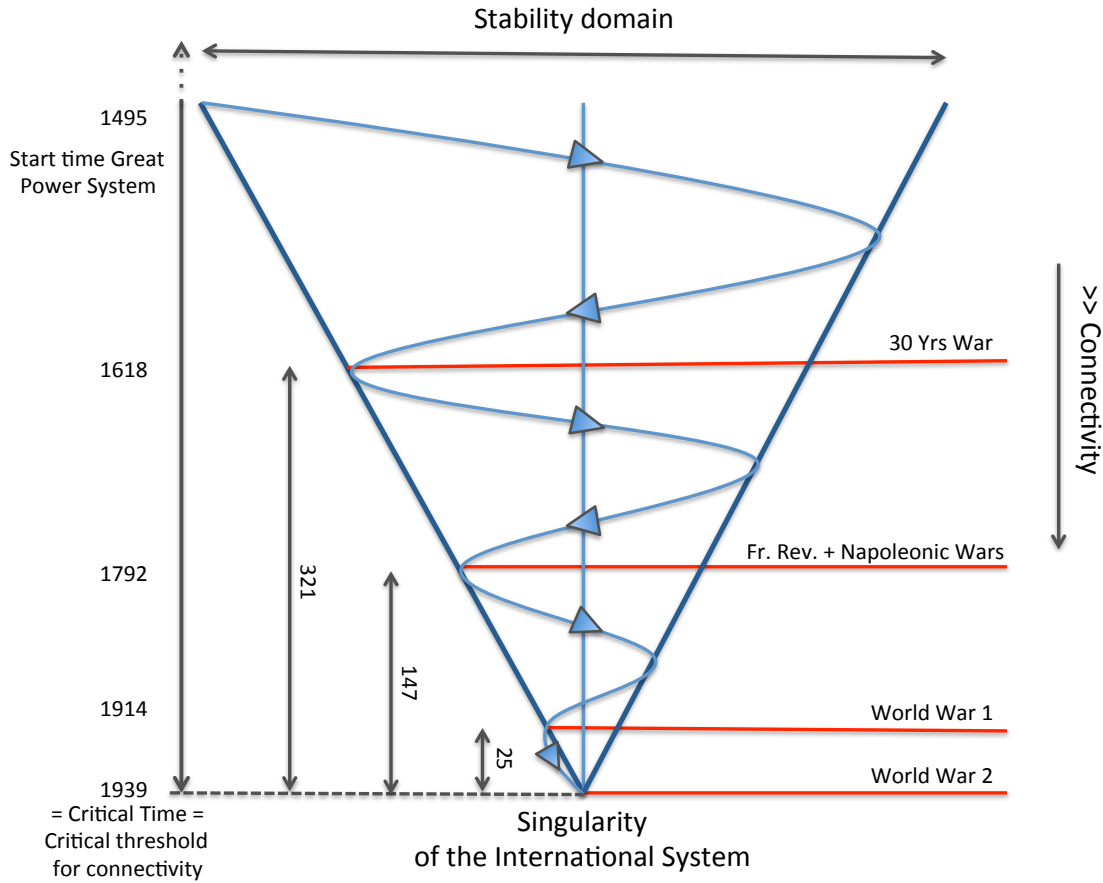

*Figure 2: This is a schematic representation of the finite-time singularity accompanied by log-periodic cycles that shaped the war dynamics and development of the International System during the period 1495 - 1945. This singularity provides us with the perspective to penetrate and decode the dynamics of the International System, and recognize a number of regularities.*

Based on the four systemic wars it is possible to distinguish four successive cycles 'delimited' by these systemic wars. Wars during the life spans of these cycles I define as 'non-systemic' wars. Historical - and systems analysis show that systemic wars typically were large scale wars, and that by means of these wars, 'new' organizing principles were introduced in the International System to improve its 'management'. Non-systemic wars did not have such impacts.

| Development of the stability and resilience of successive cycles | | | | | |
|---|---|---|---|---|---|
| | | Stability | | Resilience | |
| Cycle | Period | War frequency | GP Status dynamics | Number of Great Power wars | Life span (years) |
| 1 | 1495 – 1618 | 0.37 | 8 | 45 | 123 |
| 2 | 1648 – 1792 | 0.24 | 5 | 34 | 144 |
| 3 | 1815 – 1914 | 0.18 | 3 | 18 | 99 |
| 4 | 1918 – 1939 | 0.05 | 0 | 1 | 21 |

*Table 2. This table shows the stability and resilience of the four successive cycles; it only addresses non-systemic wars. Systemic wars are not included in this overview because they constitute a fundamentally different category. Great Power wars outside the European continent, with only one European participant, are excluded from this overview. These wars are indicative of the globalization of the International System but obscure the process of social expansion and integration in Europe (2, 2014).*

For the sake of completeness - including for the background of the calculations - I refer to "*War: Origins and Effects, How connectivity shapes the war dynamics and development of the International System*" (2, 2014).

### b. Linear increase of the stability of the International System.

I have defined the stability of the International System as the ability of the International System to sustain itself in a condition of rest, that is, in the absence of Great Power wars.  The stability of the International System can be operationalized with two measures: (1) the war frequency and (2) the status dynamics during successive cycles.

The *war frequency* of cycles can be calculated by dividing the number of non-systemic wars during the life-spans of successive cycles, with the duration of the respective life-spans.

The *status dynamics* of the International System are defined as the number of states that acquire or lose their Great Power status during the life span of successive cycles.

Based on Levy's dataset, it is possible to determine that the status dynamics decreased over time (2, 1983, p47). During the first four cycles, respectively eight, five, three, and zero status changes occurred (status changes during systemic wars are excluded).

Two of the three status changes during the third cycle concerned respectively the United States (1898) and Japan (1905). The development of this indicator over time not only emphasizes the increased stability of the European system over time, but also signals the increased impact of non-European states on the dynamics of the International System.

The war frequency *decreased* linearly: **War Frequency (t) = 0.465–1.02t, with t is the cycle number, R2 = 0,98**, which means that the stability of the International System *increased* linearly.

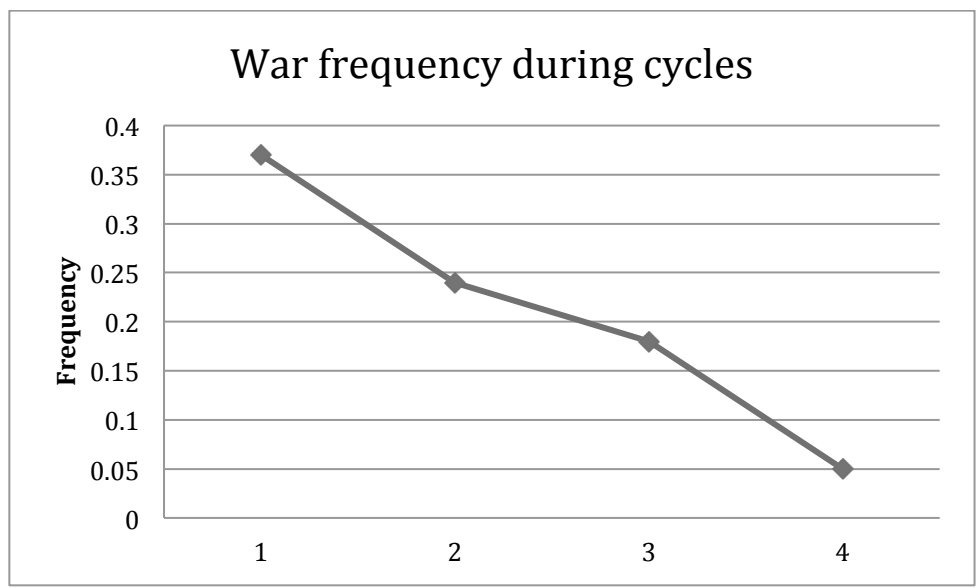

*Figure 3: This figure shows the linear decrease of the war frequency of the International System. This linear decrease in war frequency implies a linear increase in stability.*

The number of status changes during successive cycles also decreased linearly: **Status Dynamics (t) = 10.5–2.6t, with t is the cycle number, R2 = 0,99**. This is another indication for the linear increase of the stability of the International System, during the period 1495 -1945.

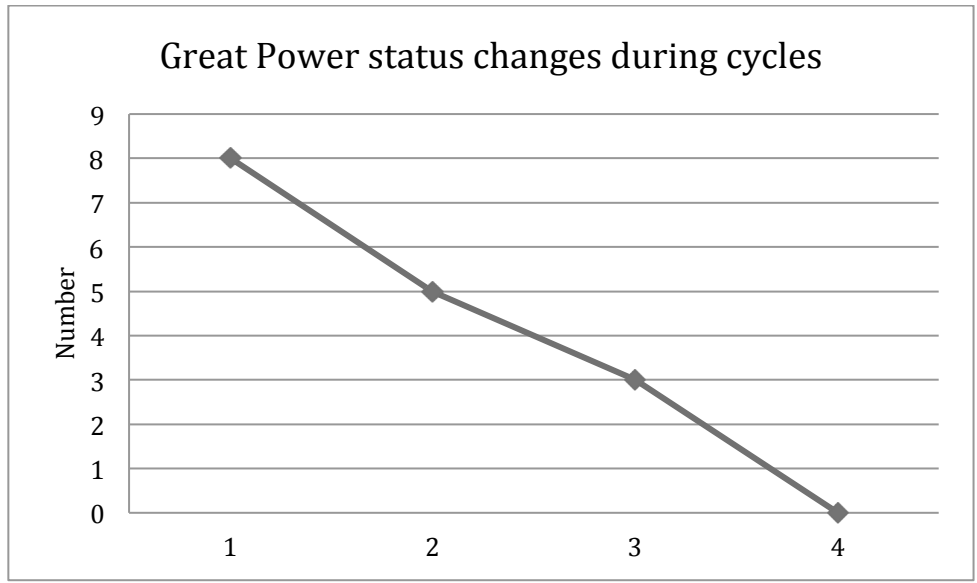

*Figure 4: This figure also shows the linear increase of the stability of the International System during four successive cycles.*

### c. Linear decrease of the resilience of the International System.

Whereas the stability of the International System *increased* linearly during the period 1495 - 1945, the resilience *decreased* linearly, at the same time: An increase of the stability went hand in hand with a decrease in resilience, this research shows. Resilience is defined as the ability of the system to sustain itself within a particular stability domain, within a certain 'cycle'.
The number of wars during the life span of a cycle is an indicator of the resilience of a cycle.
**_Resilience (t) = Number of Wars (t) = 61.5–14.8t, with t is the cycle number, R2 =0,99_**

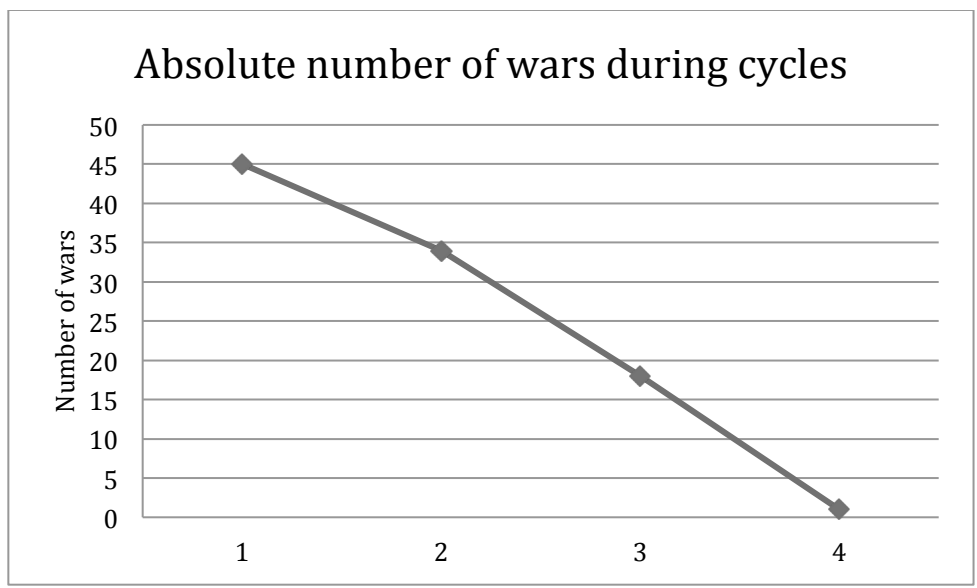

*Figure 5: This figure shows the linear decrease of the resilience of the International System.*

### d. Exponential increase of the intensity of successive systemic wars.

The intensity of successive systemic wars also shows remarkable regularities: the intensities of successive systemic wars increased exponentially. Levy defines *intensity* as "battle deaths per million European population" (1, 1983).

In this overview I did not include the fourth systemic war: the Second World War. The reason to exclude this particular systemic war is the fact that it constituted a critical transition, and cannot be 'compared' with the other systemic wars. However, the observation that the intensity grows exponentially still holds when the Second World War is also included in the analysis.

The mathematical expression for this regularity is: ***Intensity / year (t) = 5.68 e^(2.61t), with R2 = 1.00 (t = the number of the systemic war)*** and for the absolute intensity of successive systemic wars, where t also is the number of the systemic war:
***Absolute Intensity (t) = 8664.47 e^(0.62t), with R2 = 0.97.***

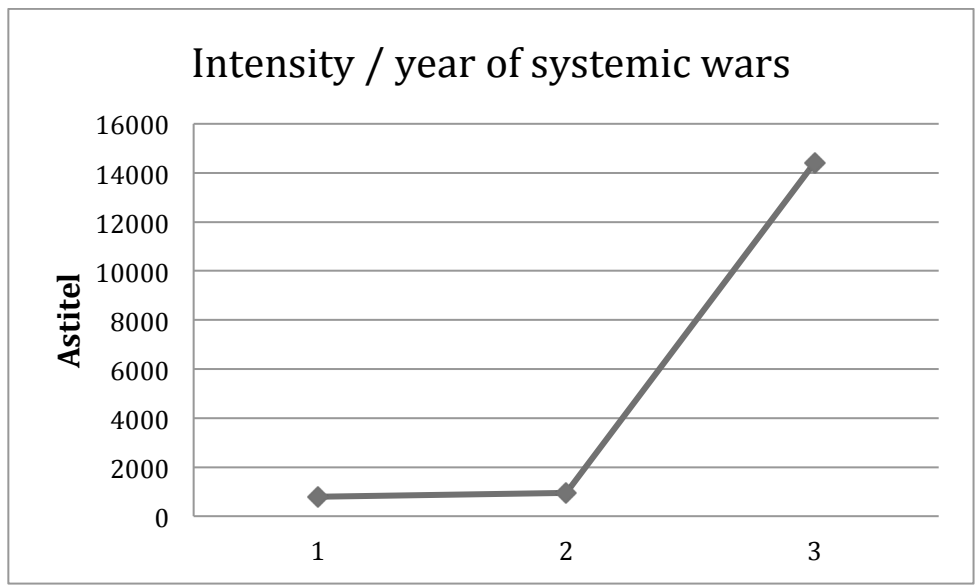

*Figure 6: Exponential increase of the ratio intensity / year of successive systemic wars.*

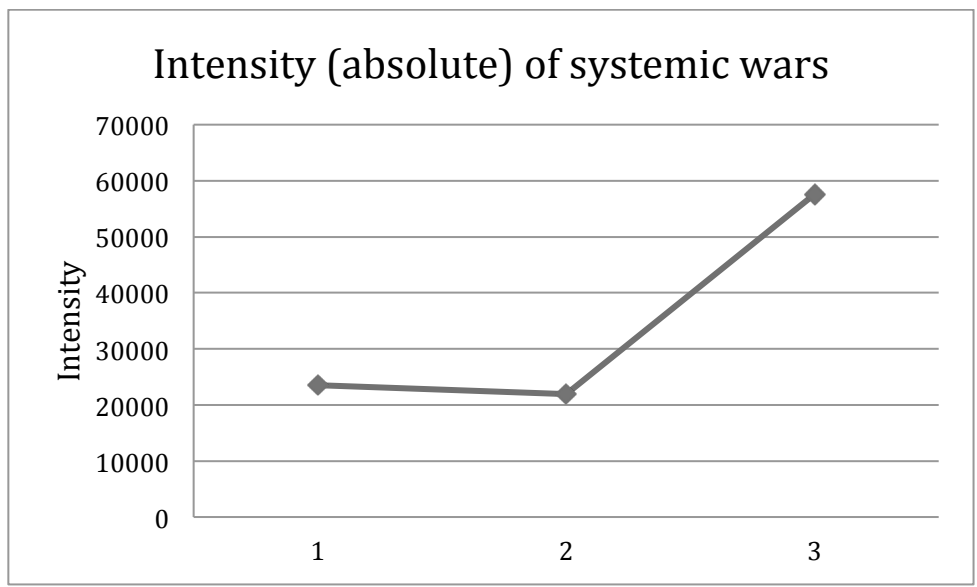

*Figure 7: Exponential increase of the absolute intensity of successive systemic wars.*

**e. Exponential decrease of the life span of successive systemic wars.**

The life span of successive systemic wars also decreased remarkably regularly. For the same reasons as mentioned above, the fourth systemic war is not included in this analysis:
**_Life span(t) = 58.1 e^(-0.61t), with R2 = 0.96._**

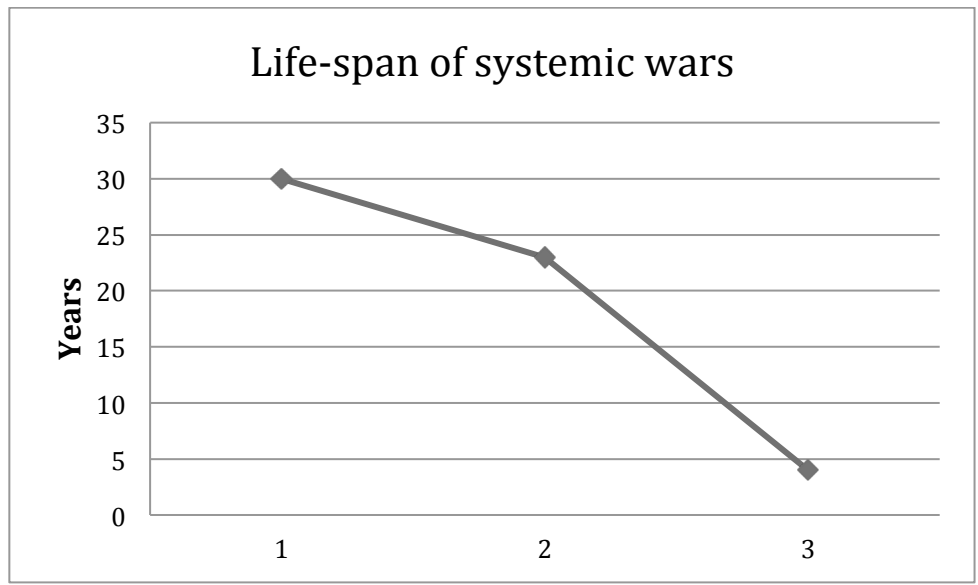

*Figure 8: The life span of the first three systemic wars. The fourth systemic war - the Second World War - is not included in this analysis, because of its function as a critical transition.*

**f. Exponential decrease of the life span of successive cycles.**

The life-span of successive cycles also decreased exponentially:
**_Life span(t) = 194.1 e^(-0.295t), with R2 = 0.92_**. It can be argued that the life spans of the first and second cycles were - or should - respectively be longer and shorter. Some historians argue that the Great Power System actually started its typical dynamics not in 1495, as Levy suggests, but somewhat earlier around 1480. For these calculations I used Levy's data.
The war dynamics during the second cycle - as I explain in the study "*War: Origins and Effects*" - were temporarily distorted, causing a lengthening of its life span. This implies that 'theoretically' the life span of the second cycle should be shorter. These corrections would produce an exponential function with an even better fit.

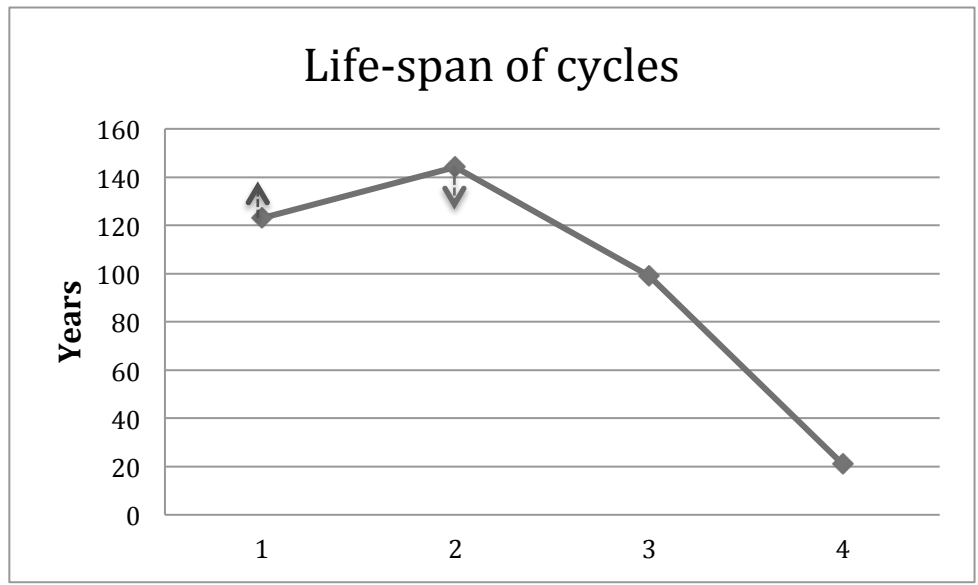

*Figure 8: The life span of successive cycles. The arrows suggest that the life span of the first and second cycle were respectively shortened and lengthened, due to specific conditions.*

Below table shows an overview of the regularities I have identified:

| | **Regularities in dynamics and development** | | | |
|---|---|---|---|---|
| | **Characteristic** | **Type of regularity** | **Mathematical formula** | **Fit (R2)** |
| 1 | Life spans of successive oscillations (singularity) | Exponential decrease | Life-span (t) = 19.6e^(0.936 t) | 0.99 |
| 2 | Stability (based on war frequencies of successive cycles) | Linear increase | War Frequency (t) = 0.465–1.02t | 0.98 |
| 3 | Stability (based on Great Power status dynamics during successive cycles) | Linear increase | Status Dynamics (t) = 10.5–2.6t | 0.99 |
| 4 | Resilience of successive cycles | Linear decrease | Resilience (t) = 61.5–14.8t | 0.99 |
| 5 | Intensity / year of successive systemic wars | Exponential increase | Intensity / year (t) = 5.68 e^(2.61t) | 1.00 |
| 6 | Absolute intensity of successive systemic wars | Exponential increase | Abs. Intensity (t) = 8664.47 e^ (0.62t) | 0.97 |
| 7 | Life-span of successive systemic wars | Exponential decrease | LS(t) = 58.1 e^(-0.61t) | 0.96 |
| 8 | Life-span of successive cycles | Exponential decrease | LS(t) = 194.1 e^(-0.295t) | 0.92 |

*Table 3: An overview of regularities identified in the dynamics and development of the International System, during the period 1495 -1945.*

## 3. <u>Statistical and causal relationships</u>

The table below shows the correlations between the variables - characteristics - discussed in the previous section.

|   |                   | 1     | 2     | 3     | 4     | 5    | 6    | 7 |
|---|-------------------|-------|-------|-------|-------|------|------|---|
| 1 | Life span of cycles |       |       |       |       |      |      |   |
| 2 | GP status changes | 0.83  |       |       |       |      |      |   |
| 3 | No. of Non-SW wars | 0.89 | 0.99  |       |       |      |      |   |
| 4 | War Frequency     | 0.82  | 1.00  | 0.98  |       |      |      |   |
| 5 | Intensity         | - 0.97| -0.92 | -0.97 | -0.91 |      |      |   |
| 6 | Intensity/Year    | -0.89 | -0.91 | -0.96 | -0.89 | 0.97 |      |   |
| 7 | Life span of SW's | 0.72  | 0.90  | 0.92  | 0.88  | 0.86 | 0.95 |   |

*Table 4. This table shows a correlation matrix for certain characteristics of the International System discussed in this report.*

This matrix shows that very strong (positive or negative) correlations exist between these characteristics. In below figure, these statistical relationships are schematically shown.

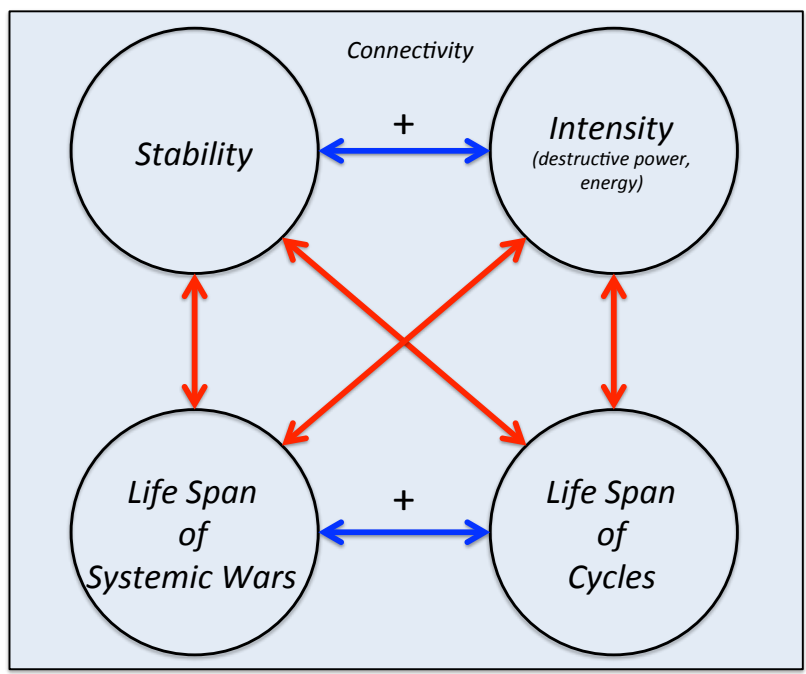

*Figure 9. This figure schematically presents the statistical relationships – correlations – between the developments of certain characteristics of the International System. Red and blue lines show strong negative and positive correlations, respectively. I assume that the causality of these characteristics lies in the connectivity of the International System. This study shows that stability, the intensity of systemic wars, the life span of systemic wars and cycles, and their development over time are closely related to the development of the connectivity of the International System over time.*

In next section I suggest that these statistical relationships can be explained with the 'unobserved variable' (factor) 'connectivity' of the International System.

## 4. Connectivity (growth)

Based on my research I assume that the connectivity of the International System grew exponentially, and that this factor - connectivity and its growth - can explain the statistical regularities and correlations that are now identified.

This assumption is based on the following reasoning: Over centuries the population grew (super) exponentially. Humans organize in 'social systems' - including in 'nation states' - to fulfill basic needs, preferably in an efficient way. Social systems also must fulfill basic requirements to ensure their survivability. Especially cooperation is an effective 'strategy' to fulfill basic needs. By cooperating it is also possible to achieve economies of scale and make more resources available to differentiate activities. Differentiation increases the change of survival.
Fulfilling basic requirements require humans and social systems - including organizations - to communicate and coordinate, in other words to connect. Cooperation and differentiation lead to increased connectivity. These dynamics - and the resulting positive feedbacks - are at the heart of a process of social expansion and integration, already underway for millennia.

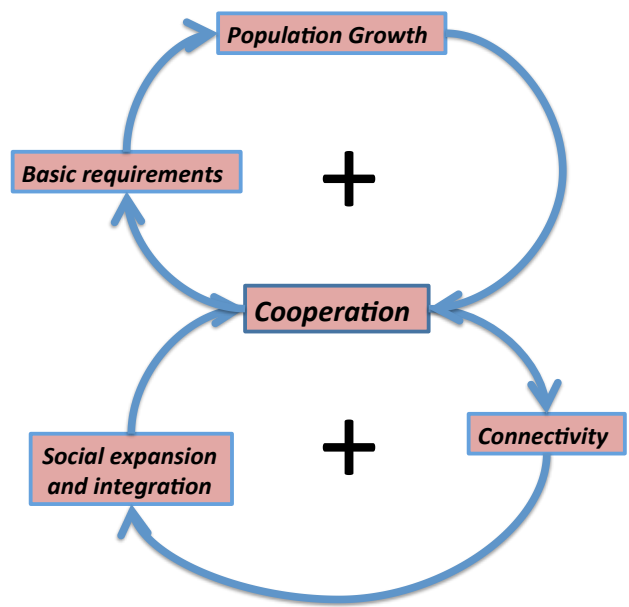

*Figure 10: This figure shows two of a number of positive feedback loops that can be identified in the dynamics between the variables discussed in this section.*

Connectivity and connectivity growth provide a consistent explanation for the observed regularities and their (very) strong statistical correlations.

Increased connectivity results in greater stability: states are more 'tightly' linked to each other, causing the war frequency to decrease. The decreased number of Great Power status changes is consistent with the increased stability of the International System: the International System over time 'settled down', it seems. Less war means less status changes. It does make sense that in an anarchistic system war frequency and status changes are causally related: wars affect status; less war implies less status changes.

As shown, a linear increase of stability goes hand in hand with a linear decrease of resilience. This relationship suggests that a more connected and more stable International System has less 'scope' for disturbances, before the need arises for the system to reorganise.

As explained, it is by means of systemic wars - which in fact are reorganisations - that new organisational design principles are introduced in the International System. These new principles are meant to better prevent new large scale wars and improve cooperation.

Periodically - and unavoidable - the International System needs to reorganise - to rebalance - its components. History shows that in an anarchistic system such a reorganisation can only be achieved by means of systemic war. A more stable system, obviously requires more 'energy' - that is systemic wars with a higher intensity - to reorganise, than a more loosely connected system.

## 5. **Some implications**

The identification of various regularities shows that the International System in fact is a deterministic system. In '*War: Origins and Effects*' I discuss numerous other indications that point to the deterministic nature of the International System.

The deterministic nature of the International System implies that we are not as much 'in control' as we often assume. For example, the (robustness of the) singularity dynamic and its strong shaping effects show that all four systemic wars would also have happenend, without the protagonists we strongly associate with these wars. Although the events leading up to these wars - and the wars themselves - are unique, the underlying dynamics are apparently not.

The fact that the conditions of the current International System are identical with the conditions of the system being studied (1495 -1945), suggests that a singularity dynamic could now well be in the making: Connectivity growth in an anachistic system still is a destructive combination.

Another implication of these research findings is that the regularities in the dynamics of the International System make it possible to develop a set of early warning signals, warning us for future systemic wars. Better information allows for better management. That is what we urgently need.